\documentstyle[prl,aps,multicol]{revtex}
\long\def\title#1{{\Large\begin{center}#1\end{center}\par}}
\long\def\address#1{\begin{center}#1\end{center}\par}
\long\def\author#1{\begin{center}#1\end{center}\par}
\def\pacs{}
\tightenlines
\begin{document}

\draft

\title{The gas of elastic quantum strings in 2+1 Dimensions: finite temperatures.}



\author{Sergei I. Mukhin}
\address{Moscow Institute for Steel and Alloys,
Theoretical Physics Department, Leninskii pr. 4, 117936 Moscow,
Russia}      

\author{Wim van Saarloos and Jan Zaanen}
\address{Instituut-Lorentz for Theoretical Physics, Leiden University,
P.O. Box 9506, 2300 RA Leiden, The Netherlands}

\vspace{3mm}
\date{today}
\maketitle

\begin{abstract}
Motivated by the stripes of the high Tc cuprates the problem was introduced recently
of a system of free elastic quantum strings interacting via a hard-core condition 
embedded in 2+1
dimensions. At zero-temperature this system is always a solid due to `quantum-entropic'
 (or `kinetic')
interactions which dominate at long wavelength. The high temperature limit of this 
problem corresponds
with thermally meandering elastic lines in 2D and this system is well known to be 
dominated by
litteral entropic interactions. Here we analyze in detail what happens in between 
zero- and high
temperature. We identify a `renormalized classical' intermediate regime where the 
on-string fluctuations
have become predominantly quantum-mechanical. Surprisingly, the entropic interactions 
keep their 
high temperature nature in this regime. At a low, but finite temperature the 
quantum-mechanical 
kinetic interactions take over rather suddenly. Despite their origin in long wavelength
 quantum fluctuations
these are not affected by thermal fluctuations when temperature is low enough.  
 
\end{abstract}

\section{Introduction}

The focus of quantum condensed matter has been traditionally on systems of
particle-like excitations. Recently the evidence has been growing that at
least in the strongly interacting electron systems as realized in correlated oxides
novel types of quantum-mechanical self-organizations are taking place on
mesoscopic time- and length scales. Although these are fluctuating textures,
there exists substantial emperical evidence that they bear a direct relationship
with the static stripe-phases. 

The stripe phase is a novel form of electronic order which is found in doped 
Mott-insulating 
oxides \cite{tran}. Because of topological reasons\cite{geom}, the carriers bind to line-like
textures in the 2+1D cuprates and nickelates, and these `rivers of charge' are
separated by Mott-insulating magnetic domains. The static stripes are reasonably 
well understood  as generic instabilities of the doped Mott-insulating state
\cite{str}. However,
experimental results suggest that on intermediate
scales (nanometer length scales, $\sim 10$ meV energy scales) the electron system 
in the cuprate superconductors tends to stripe order\cite{emery}, to flow away to the anomalous 
physics
of high $T_c$ superconductivity at larger distances and longer times. 
It is often speculated that the high Tc superconducting state is in one or the other 
way
related to this quantum disordered stripe state (the `dynamical stripes').

On the theoretical side, the obvious problem is that little is known in general
about a `complex' quantum state of matter of the kind
suggested by the dynamical stripe phenomenology.  This motivated us to look into the
following elementary question: what can be said in general on the physics of
a gas of quantum fluctuating elastic lines in two embedding space dimensions? These 
lines can be alternatively called strings \cite{zahovs,eskes,morrais,nayak,kivel,hassel} and 
to have a meaningful definition
of a gas it is natural to let these strings merely interact via an excluded
volume (or non-crossing) condition \cite{finT}.

One of us recently suggested an answer \cite{jan}: due
to an order-out-of disorder mechanism, the ultimate faith of this gas is that it 
solidifies always
at zero temperature. This can be either viewed as
a 2+1D extension of the mechanism responsible for the algebraic order in the
1+1D Luttinger liquid, or as a quantum version of the classical entropic
repulsions familiar from the statistical physics of lines and membranes \cite{copper}.
 A central result of reference \cite{jan} is that this string-solid 
should be characterized by a long-wavelength compression modulus which
depends on the average string separation $d$ in a stretch exponential fashion,
$\sim exp(-A d^{\alpha})$ with $\alpha \sim 2/3$. This argument was based on
an elegant, but non-exact self-consistent phonon method introduced quite
some time ago by Helfrich \cite{helfrich} in the context of biological membranes. 
The stretched
exponential turns out to reflect a highly untrivial and counterintuitive phenomenon.
Usually entropic interactions emerge from short distance physics. The essence of
the mechanism is that entropy is paid at collisions because of the hard-core
condition, and collisions occur at short distances. For the case of elastic
quantum strings this is qualitatively different. A single string shows algebraic
order and for this reason its long-wavelength fluctuations are the most dangerous ones.
According to the Helfrich method, these long-wavelength on-string fluctuations 
are responsible for the induced modulus\cite{jan}. Very recently, this result was confirmed in a
numerical simulation by Nishiyama \cite{nishi}, finding $\alpha \simeq 0.8$ instead 
of $\alpha=2$ as would follow from the argument based on simple collisions. 

The main focus of this paper is on the {\em finite} temperature physics of this quantum
string gas. This finite temperature behavior is in itself rather untrivial. At zero
temperature the rigidity in the system is driven by a gain of {\em kinetic} energy
associated with long wavelength on-string fluctuations. At the same time, the high
temperature, fully classical limit corresponds with the problem of elastic lines
meandering over the 2D plane due to thermal fluctuations. This is a classic statistical
physics problem which is fully understood \cite{zahovs,copper,pot}. It is also characterized by a net 
coarse-grained modulus, now proportional to the square of temperature, originating 
in litteral entropic interactions.
The question is, how to connect the zero-temperature limit with the high temperature 
limit? Since the order at zero temperature is driven by long wavelength fluctuations 
one could
be tempted to argue that at any finite temperature the entropic interactions should 
take over -- at sufficiently long wavelength, thermal fluctuations usually overwhelm 
quantum
fluctuations for any finite temperature. If this would be the correct answer there 
could be a
potential problem with the argument of \cite{jan}. To show that the string gas solidifies
 at zero
temperatures it has to be demonstrated that dislocations cannot proliferate. It is well
 known that
in the classical (high temperature limit) dislocations always proliferate. 
The effective elastic constant $\sim T$ and under this condition the criterium for
the Kosterlitz-Thouless (KT) transition (binding of dislocations) is never 
satisfied\cite{zahovs,copper,pot}.  However, upon
adding any tension of non-entropic origin the KT transition immediately shifts to a 
finite temperature
and the zero temperature state is protected against free dislocations. One of us
\cite{jan} asserted 
that the
induced modulus is non-zero at zero temperature due to the quantum fluctuations and 
this modulus
can therefore serve the purpose of protecting the crystallinity of the ground state.

Here we address these matters by analyzing the finite temperature regime explicitely 
using the Helfrich method. Starting from the high  temperature,
classical regime one first enters a `renormalized classical' regime where the 
quantum-mechanical
nature of the on-string fluctuations becomes noticable because temperature is lower 
than the
characteristic Debye temperature of the on-string modes. Naively one would expect 
either a
qualitative change in the nature of the entropic interactions as compared to the high
temperature limit, or at the least a quantitative change in the sense that numerical 
factors
in the classical result for the induced modulus become different. We were quite 
surprised
to find that in fact the induced modulus is not changed at all. As we will explain, the
 reason
turns out to be that the fluctuations driving the entropic interactions occur at 
frequencies
which are low as compared to the quantum UV cut-off. Upon further lowering temperature,
 suddenly
the quantum-kinetic interactions take over, at a temperature scale associated with the 
zero-temperature
modulus. This quantum modulus protects itself: despite its origin in long wavelength
 on-string fluctuations,
it carries an associated energy scale and when temperature becomes lower than this 
scale the thermal
fluctuations freeze out. As a result, in this low temperature regime the thermal 
contributions appear as 
corrections to the zero temperature modulus.
    
The plan of the paper is as follows. In section II the string gas model is specified,
and  the dimensionless parameters governing 
the crossovers between the different temperature regimes are estimated using the
simple collision picture. In section III the Helfrich method \cite{helfrich}
is introduced and used to
refine the  estimates for the characteristic scales of section II. In section IV the
behavior of the induced modulus in the various regimes are derived  and discussed in
detail. In the concluding section we put our findings in perspective.   

\section{Characteristic scales: the collision picture.}

One can acquire some insight regarding the origin of entropic repulsions in terms
of a simple physical picture. The basic idea is that the worldlines (particles) or 
worldsheets (strings)
once in a while collide when they meander in space-time. Entropy (high temperature) 
or kinetic 
energy (zero temperature)  is paid at these collisions because of the 
non-crossing condition. This raises the (free) energy, and this energy increase 
translates into repulsive interactions at longer distances. Although this argument
turns out to be not quite right for the zero-temperature string gas, it is 
qualitatively
correct at high temperatures. In this section we will use this argument to obtain a
crude picture of the physics at all temperatures, which will be refined in the next
sections using Helfrich's self-consistent phonon method.  

The string gas is defined as a system of non-intersecting elastic quantum strings 
embedded in 2+1D 
space-time. In path-integral language this corresponds with the statistical physics of
a system of non-intersecting, elastic `worldsheets'.  
In addition, since it follows self-consistently from the theory that dislocations do
not proliferate at zero temperature the strings can be considered to be directed along 
the $y$-axis 
from the on-set. The transversal displacements of the strings are parametrized  in 
terms of a field 
$\phi_{i}(y,\tau)$ describing the motion of the $i$-th string in the $x$ direction 
($\tau$ is imaginary time), and supplemented by the avoidance condition,

\begin{eqnarray}   
\phi_{1}(y,\tau)<\phi_{2}(y,\tau)< \ldots < \phi_{N}(y,\tau)\;.
\label{hardcore}
\end{eqnarray}

\noindent
The partition function written in the  form of a functional integral is

\begin{eqnarray}
&&Z=\displaystyle\prod_{i=1}^{N}\prod_{y,\tau}\int\;d\phi_i(y,\tau)\exp{\{-
S/\hbar\}}
\;, \nonumber\\
&&S=\displaystyle\frac{1}{2}\int_0^{\frac{1}{T}}\;d\tau\int\;dy\sum_i\left[\rho_c
(\partial_\tau\phi_i)^2+\Sigma_c(\partial_y\phi_i)^2\right]\;,
\label{partition}
\end{eqnarray}

\noindent
where the temperature T is expressed in energy units ($\hbar = k_B = 1$). In Eq. 
(\ref{partition}) $\rho_c$ is
the linear mass density and $\Sigma_c$ the string tension, such that 
$c=\sqrt{\Sigma_c/\rho_c}$ is the on-string sound velocity. As in \cite{jan}, a  
lattice
regularization with lattice constant $a$ is chosen, such that the average interstring
distance is $d=a/n$, where $n$ is the density of strings. The
ultraviolet (UV) momenta and frequency cut-offs on a single string (i.e. for the 
intrastring vibrations) are therefore $Q_s = \pi/a$ and $\omega_s=cQ_s$, respectively.
Throughout this paper we use as a convention a subscript $s$ when a quantity relates 
to a single string and a subscript $g$ when it relates to the system (`gas') of strings.

Let us now turn to the collision-argument.  Since the strings only interact via a 
non-intersection 
condition it is obvious that at sufficiently short times and distances the strings
behave like a non-interacting system because collisions do not occur. A characteristic
scale (collision length) can be identified where the probability of a collision becomes
 of order unity and
at larger scales the physics is set by the collisions. At every collision an entropy 
$\sim k_B$
(at high $T$) or a kinetic energy $\sim \hbar$ (at $T=0$) is paid and an estimate 
follows
for the induced modulus by dividing this characteristic (free) energy cost by the 
characteristic
distance ($d$). Hence, all what has to be calculated is the collision length and this 
can be
determined by   evaluating the  mean-square of the single string
meandering amplitude  $\langle(\phi(y)-\phi(0))^2\rangle$ as a
function of a distance $y$ between two points on a string. Equating this 
quantity with the square of the average interstring distance $d$ one obtains
a characteristic distance $y = l_c$ (and time $\tau_c = l_c / c$) where the
probability for a collision to occur becomes of order unity. 

After fourier transforming  the  single string action we arrive at the standard result,

\begin{eqnarray}
&& \langle(\phi(y)-\phi(0))^2\rangle \sim \int \langle |\phi_k| ^2\rangle(1-\cos{(ky)})
d\,k
=\int \frac{\hbar}{\rho_c c k}({1}/{2}+n_k)(1-\cos{(ky)})d\,k\nonumber\\
&& \sim \int_{\pi/y}^{\pi/a} \frac{\hbar}{\rho_c ck}\displaystyle [{1}/{2}+
(\displaystyle \exp[{\hbar ck}/{T}]-1)^{-1}]d\,k
\label{meander}
\end{eqnarray}
Consider first Eq. (\ref{meander}) in the high-temperature (classical) limit,
\begin{eqnarray}
T\gg \frac{\pi\hbar c}{a}\; .
\label{hight}
\end{eqnarray}

\noindent
The mean-square meandering amplitude becomes 

\begin{eqnarray}
&& \langle(\phi(y)-\phi(0))^2\rangle \sim  \int_{\pi/y}^{\pi/a} 
\frac{\hbar}{\rho_c ck}
\displaystyle [{1}/{2}+(\displaystyle \exp[{\hbar ck}/{T}]-1)^{-1}]d\,k\nonumber\\
&&\approx \int_{\pi/y}^{\pi/a} \frac{\hbar}{\rho_c ck}\displaystyle [{1}/{2}-1/2+
T/(\hbar ck)]d\,
k\sim \frac{Ty}{\Sigma_c}
\label{cmeander}
\end{eqnarray}

\noindent
Equating this to $d^2$ yields for the collision length $l_c$ the well-known result\cite{copper,zahovs,jan},

\begin{eqnarray}
{Tl_c}/{\Sigma_c}\approx d^2\,.
\label{lhigh}
\end{eqnarray}

\noindent
In the classical context, the induced modulus follows directly by dividing the 
free-energy 
cost
associated with the collisions $\sim T S \sim T \times (1/l_c) \times k_B$ ($1/l_c$ is 
the
collision density) by the  characteristic length $d$ (interstring
separation). Thus, Eqs. (\ref{cmeander}) and (\ref{lhigh}) lead to an estimate:
$B\propto T^2/d^3$. This shows the correct dependence on temperature and density.
Alternatively, this directed 2D string gas can also be
interpreted as a system of  1+1D hard-core bosons which is in turn equivalent to
non-interacting spinless fermion gas.  In the latter,  the
collision length corresponds with the Fermi-energy and using these trivial 
spinless-fermion
results one finds  that the estimate for the high temperature limit of the string gas 
is
qualitatively correct\cite{jan}.  

The high temperature limit is defined as usual as the regime where temperature is 
larger
than the highest phonon frequency which is in our case clearly corresponding with the
UV cut-off of the on-string modes ($Q_s, \omega_s$). Hence, upon lowering temperature 
one will reach
a on-string Debye temperature below which the quantum-mechanical nature of the modes
on the string will become noticable. This cut-off frequency is $\omega_s = (\pi / a)
\sqrt{\Sigma} / \sqrt{\rho}$, defining a dimensionless ratio ($\hbar = k_B = 1$),

\begin{equation}
\nu_s = { {\omega_s} \over T } = { {\pi \sqrt{\Sigma}} \over { a \rho T} }
\label{nustr}
\end{equation} 

Therefore, when $\nu_s > 1$ one enters a regime where the high frequency on-string 
modes
are freezing out, and we call this the renormalized classical regime. Since the 
on-string
modes drive the meandering of the string one would expect that something happens with
the entropic interactions around $\nu_s = 1$. As it turns out, however, the induced 
modulus continues to show its high temperature behavior while even the prefactors are
not affected. This counter-intuitive fact might be understood qualitatively 
on the basis of Eqs. (\ref{cmeander})  and (\ref{lhigh}). 
Let us estimate the characteristic energy of the on-string mode,
which has a wave-length of order of the collision length  $l_c$:  $\omega_{coll.} = 
\hbar
c/l_c\sim \hbar cT/d^2\Sigma_c$. This implies that this energy scales down linearly
with the temperature $T$ while it corresponds with the {\em lower bound} in energy on
the modes which contribute to the collision length. Modes with lower energy have a 
wavelength
larger than the collision length and do therefore not contribute. Hence, instead of 
having
all modes contributing to the thermal fluctuations, for $\nu_s < 1 $ only modes with 
frequencies
$\omega_{coll} = \alpha_{coll} T < \omega < T$ contribute. At first sight one would 
expect this to give 
rise to a gross change. However, on closer inspection one finds it to be more subtle.
A first requirement is that $\alpha_{coll} << 1$, otherwise the contribution of the 
thermal fluctuations
would be completely quenched out. To estimate $\alpha_{coll}$ we recall that the 
importance of quantum
fluctuations in the string gas is measured by the dimensionless ratio,

\begin{equation}
\mu_g = { {\hbar} \over { \rho_c c d^2} },
\label{boer}
\end{equation}

\noindent
corresponding with the ratio of $\hbar$ with the dimensionful quantities characterizing
 the problem:
the coupling constant of the gas or the ``de Boer parameter''\cite{zahovs}. 
$\mu_{g}$ has to be 
small 
compared
to unity because otherwise quantum fluctuations become important at the lattice scale, 
and the
continuum description fails. Using Eq. (\ref{lhigh}) we find immediately that 
$\alpha_{coll} = \omega_{coll} / T 
\sim \mu_g << 1$! Hence, the window in the modespectrum contributing to the thermal 
meandering is in this
sense large and it remains large regardless the fact that temperature becomes less 
than $\omega_s$, 
at least initially. 

It is still so that the short-wavelength modes with frequency $\omega > T$  are no 
longer available for
the thermal meandering. However, a simple calculation shows that their contribution 
is exponentially small.
Consider again  Eq. (\ref{meander}), realizing that only the zero-point contribution 
matters for momenta
in the interval $T/(\hbar c) < k < \pi / a$ while the thermal contribution dominates 
in the interval
$\pi / l_c < k < T / (\hbar c)$. Carrying out the integrals,

\begin{eqnarray}
&& \langle(\phi(y)-\phi(0))^2\rangle \approx  {1}/{2}\int_{T/\hbar c}^{\pi/a} 
\frac{\hbar}{\rho_c ck}d\,k + \int_{\pi/y}^{T/\hbar c} \frac{T}{\rho_c
(ck)^{2}} d\,k\nonumber\\
&&\approx \displaystyle \frac{\hbar}{2\rho_c c}\displaystyle\ln{\frac{\pi \hbar c}{aT}}
+
\frac{Ty}{\pi\Sigma_c}+O(\mu d^2)\,.
\label{rcmeander}
\end{eqnarray}

\noindent
The logarithmic term in Eq. (\ref{rcmeander}) is due to the quantum zero-point string
meandering coming from the modes with $\omega > T$ while the second term corresponds 
with the thermal meandering. Equating expression in (\ref{rcmeander}) to $d^2$ we find 
a "renormalized
classical" result (\ref{lhigh}):

\begin{eqnarray}
{Tl_c}/{\Sigma_c}\approx d^2(1+O(\mu))\,,
\label{lrhigh}
\end{eqnarray}

\noindent
provided that the logarithmic term is small. For this to be true, the temperature 
should
be above a classical-to-quantum crossover temperature T$_0 << \omega_s$. At 
sufficiently low 
temperature eventually the logarithm will dominate and we can therefore estimate 
T$_0$
by neglecting the second term in Eq. (\ref{rcmeander}) and equating the logarithm to
$d^2$. It follows that,

\begin{eqnarray}
T_0 \sim \displaystyle\frac{\pi \hbar c}{a}\exp{\left[-\frac{2}{\mu}\right]}
\label{tcross11}
\end{eqnarray}

\noindent
As we shall see in the following sections, this estimate for $T_0$ is actually
flawed, while the correct answer is the stretched exponential following from
the Helfrich self-consistent phonon method. However, for the purpose of crude 
estimations it suffices. T$_0$ is small compared to the on-string Debye temperature,
because $\mu_g < 1$ and as long as $T >> T_0$ the logarithm can be neglected. 
As long as this is the case nothing changes as compared to the high temperature limit,
because Eq.'s (\ref{lrhigh}) and (\ref{lhigh}) are just the same! Hence, we arrive
at the counter-intuitive result that the collision density and thereby the induced
modulus is insensitive to the freezing out of the on-string modes in the renormalized
classical region in between $T_0$ and $\omega_s$.

Finally, at temperatures $T < T_0$ quantum-mechanics is expected to take over. 
The wavelength
of the thermally excited modes are now large as compared to the collision length set 
by the
quantum fluctuations. Therefore, we can neglect the Bose factor in Eq. (\ref{meander}) 
completely,

\begin{eqnarray}
&& \langle(\phi(y)-\phi(0))^2\rangle \sim  \int_{\pi/y}^{\pi/a} \frac{\hbar}{\rho_c ck}
\frac{1}{2}d\,k=
\frac{\hbar}{2\rho_c c}\ln{\left(\frac{y}{a}\right)}
\label{qmeander}
\end{eqnarray}

\noindent
Equating the latter estimate to $d^{2}$ we find an exponentially large , though finite 
value
of the collision length $l_c$:

\begin{eqnarray}
l_c\sim a\exp{\left[\frac{2}{\mu}\right]}
\label{qlc}
\end{eqnarray}

\noindent
Thus, an effective quantum-meandering induced repulsion is expected to be
exponentially small but finite in the zero-temperature limit. We repeat,
this last estimate is flawed as the comparison with numerical results
shows. In any other regard we will find that the simple estimates presented
in this section are consistent with the results of the more involved 
Helfrich method. This includes the basic observation that the zero-temperature
induced modulus survives at small but finite temperatures
$T < T_0$. This finding suffices to protect the solidification of the string
gas at zero temperature because the dislocations can only proliferate at
a Kosterlitz-Thouless transition at a temperature $T_{KT} \sim T_0$, 
 according to the exact result of
Pokrovsky and Talapov \cite{pot}.  


\section{Characteristic scales according to Helfrich's method.}

As explained in \cite{jan}, within the Helfrich scheme \cite{helfrich} the avoidance 
condition Eq. (\ref{hardcore}) is dropped, and a self-consistently determined 
effective rigidity $B$ is introduced instead. The parameter $B$ governs the
interstring  interactions in the effective long wave-length action, $S_{eff}$,
which substitutes for the exact action $S$ as defined in Eq. (\ref{partition}),

\begin{eqnarray}
S_{eff}=\displaystyle\frac{1}{2}\int\;d\tau\;dx\;dy\left[\rho(\partial_\tau\psi)^2+
\Sigma(\partial_y\psi)^2 + B(\partial_x\psi)^2\right]\;.
\label{action}
\end{eqnarray}

\noindent 
Here $\psi(x,y,\tau)$ is the coarse grained, long-wavelength displacement field
taking over from the bare field $\phi_i(y,\tau)$, and $\Sigma=\Sigma_c/d$, 
$\rho=\rho_c/d$. The 
self-consistency equation from which the induced rigidity $B$ can  be determined is 
nothing else
than the well-known relation between the free energy $F$ of an elastic medium and its 
compression
modulus, 

\begin{eqnarray}
B=\displaystyle d^2\frac{\partial^2(\Delta F(B)/V)}{\partial d^2}\;.
\label{b}
\end{eqnarray}

\noindent 
Here $V$ is the system's real-space volume, i.e. in the case of  two-dimensional
space the area : $V\rightarrow L^2$, where $L$ is the linear dimension of this space. 
The free energy difference,  $\Delta F(B)$, is defined as,:

\begin{eqnarray}
\Delta F(B)\equiv F(B)-F(B=0)=-T\ln{\{Z(B)/Z(B=0)\}}\;.
\label{selfb}
\end{eqnarray}

\noindent
$\Delta F(B)$ is most conveniently derived using the standard procedure, representing 
the free field as a set of non-interacting quantum harmonic oscillators with  
frequencies,

\begin{eqnarray}
\omega_q=\sqrt{\Sigma q_y^{2}+Bq_x^2}\;,
\label{omegaq}
\end{eqnarray}

\noindent
defined for wave-vectors $\vec{q}$ in momentum space bound by UV cut-off's: 
$|q_y|\leq\pi/a\equiv Q_s$; 
$|q_x|\leq \pi/d\eta\equiv Q_g$, where $Q_s$ and $Q_g$ are the cut-off's along- and 
perpendicular to  the string direction, respectively. $\eta\sim 1$ is a "fudge-factor",
correcting for inaccuracies coming from short distance mode-coupling effects. 
\cite{jan}. Using this representation for $S_{eff}$, one easily obtains familiar
expressions for the partition function $Z$ and the free energy $F$ of an ideal gas of 
harmonic oscillators \cite{landau},

\begin{eqnarray}
&&Z=\prod_{q}\exp{(-\beta\hbar\omega_{q}/2)}
(1-\exp{\{-\beta\hbar\omega_{q}\}})^{-1}\;,\nonumber\\
&&F=-T\ln{Z}=\sum_{q}\frac{\hbar\omega_{q}}{2} +k_{B}T\sum_{q}
\ln{(1-\exp{\{-\beta\hbar\omega_{q}\}})}=F_{0} + F_{T}\;,
\label{freeenergy}
\end{eqnarray}

\noindent
where $\beta\equiv 1/T$.
In the thermodynamic limit  $L Q_s\gg 1, \; LQ_g\gg 1$, and the momentum 
summation in Eq. (\ref{freeenergy})
can be substituted with an integration, leading to the following expression for 
$\Delta F(B)$ ($\hbar=1$),

\begin{eqnarray}
&&\Delta F(B)/L^2 = \displaystyle
\frac{1}{(2\pi)^{2}}\int_{-Q_{s}}^{Q_{s}}
d q_{y} \int_{-Q_{g}}^{Q_{g}}d q_{x}\;\left\{\frac{\sqrt{\Sigma q_{y}^{2}+
B q_{x}^{2}}}{2\sqrt{\rho}}-\frac{\sqrt{\Sigma
q_{y}^{2}}}{2\sqrt{\rho}}\right\} + \displaystyle
\frac{T^{3}\rho}{{\pi}^{2}\sqrt{\Sigma B}}I (\nu_s,\nu_g)= \nonumber\\ 
&&=\frac{T^{3}\rho}{{\pi}^{2}\sqrt{\Sigma B}}\left[\displaystyle\int_{0}^{\nu_s}
d y \int_{0}^{\nu_g}d x\;{\left(\frac{\sqrt{x^2+y^2}}{2}-\frac{y}{2}\right)}+
\displaystyle I (\nu_s,\nu_g)\right]\equiv \Delta_0 + \Delta_T 
\label{deltaf}
\end{eqnarray}


\noindent 
The first term corresponds with the zero-temperature quantum contribution to 
the free
energy ($\Delta_0$), while the second term ($\Delta_T$) contains the thermal 
contributions, proportional to,  

\begin{eqnarray}
I(\nu_s,\nu_g)=\displaystyle\int_{0}^{\nu_s}
d y \int_{0}^{\nu_g}d x\;\ln{\left(\frac{1-\exp{\{-\sqrt{x^2 +
y^2}\}}}{1-\exp{\{-y\}}} \right)}\;.
\label{I}
\end{eqnarray}

\noindent 
Besides the prefactor in front of $I$, temperature enters only through the upper bounds
 in the integrals, 
Eq. (\ref{I}).
$\nu_s$ we encountered already in the previous section as the dimensionless 
on-string Debye temperature Eq. (\ref{nustr}). 
The novelty is that we have also to introduce a dimensionless ratio associated with the
 Debye temperature for
the fluctuations perpendicular to the strings:

\begin{equation}
\nu_g = (\pi/\eta d)\sqrt{B}/(T\sqrt{\rho}) = \omega_g/T
\label{nugas}
\end{equation}

\noindent
where  $\omega_g$ plays the role
of frequency cut-off (Debye temperature) for the interstring vibrations along
the $x$ axis. This cut-off has to be determined self-consistently because it
obviously depends on the induced rigidity itself. This is a complicating factor.
This $\omega_g$ is of crucial importance for balancing the relative importance
of thermal- and quantum fluctations, while at the same time it is itself determined
by this balance. We will demonstrate later that $\omega_g$ in fact acts 
according to the naive expectations. When $T$ becomes less than $\omega_g$ the 
frequency
window available for thermal fluctuations quickly diminishes and the quantum 
fluctuations
take over completely already at a finite temperature. All what remains is to 
calculate
self-consistently what $\omega_g$ is and this will turn out to be proportional 
to the 
zero temperature $B$ for any temperature such that $\nu_g > 1$.

Another matter is the renormalized classical regime, defined by $\omega_g < T < 
\omega_s$.
We will find that the self-consistent phonons do confirm the arguments of the previous
section; the induced modulus is highly insensitive to the difference between the 
quantum-mechanical
versus classical nature of the `high' frequency phonons.   

To complete the analysis of the dimensionless parameters characterizing this problem, 
we notice that
there are actually two coupling constants. The first one arises already on the level 
of the bare strings,
and it parametrizes the importance of quantum physics for an isolated string,

\begin{eqnarray}
\mu_s=\hbar/(\rho_c c a^2)\;, 
\label{mu0}
\end{eqnarray}

\noindent
and we notice that this is related to the coupling constant of the string gas, 
Eq. (\ref{boer}), through,

\begin{eqnarray}
\mu_g \equiv\mu_s a^2/d^2 =\hbar/(\rho_c c d^2)
\label{dimdef}
\end{eqnarray}
 
It now becomes directly clear why $\mu_g << 1$; the ratio $(a/d)^2 < 1$ while the
 meaning of $\mu_s =1$ is
that the quantum fluctuations become strong on the scale of the lattice constant 
thereby invalidating the
notion of a single continuum  string. 

Interestingly, the single string parameter $\mu_s$ arises naturally in the Helfrich 
method
when the self-consistency equation (\ref{b}) for the induced
rigidity $B(d)$ is written in  dimensionless form:

\begin{eqnarray}
z=\mu_s \displaystyle\frac{\partial}{\partial
s}\left(s^2\frac{\partial\Delta(z,s,\zeta)}{\partial s}
\right)\;,
\label{bdim}
\end{eqnarray}

\noindent
where the dimensionless variables are defined as follows:

\begin{eqnarray}
s=\frac{Q_g}{Q_s}=\frac{a}{\eta d},\; z=s\frac{B}{\Sigma},\; \Delta(z,s,\zeta)=
\frac{a^2}{\hbar\omega_s}\frac{\Delta F}{L^2}\;. 
\label{dimdef0}
\end{eqnarray}

This demonstrates that $\mu_s$ is the quantity setting the overall scale of the 
problem -- everything
else follows from the self-consistency condition. Summarizing, we have established 
that the string
gas problem is characterized by four dimensionless quantities. Besides the coupling 
constants $\mu_s$
and $\mu_g$, governing the zero-temperature physics, we also find the two Debye 
temperatures $\nu_s$
and $\nu_g$ governing the balance between  quantum- and thermal fluctuations. 
There are alltogether
three regimes: (a) $\nu_g > 1$, the low temperature, quantum dominated regime, (b) 
$\nu_s < 1$, the
high temperature regime dominated by  thermal fluctuations, (c) $\nu_g < 1 < \nu_s$, 
the renormalized
classical regime where the dynamics is quantum-mechanical at short distances while the 
system as a whole
still behaves as if it is in the high temperature limit.

\section{Induced moduli from low to high temperature.}

In this section we will analyze in detail the behaviors of the induced moduli in the 
various
temperature regimes as follow from the Helfrich method. In the first two subsections 
we will
revisit the zero-temperature and high temperature regimes which were already analyzed 
by one
of us\cite{jan}. In the last two sections we approach the intermediate temperature renormalized 
classical
regime both from the high temperature- and low temperature side, demonstrating that 
(a) the
induced modulus does not change upon entering the renormalized classical regime from 
the high
temperature side, while (b) $\nu_g$ is
actually finite, with the implication that the zero-temperature quantum rigidity 
already takes
over at a small but finite temperature.

\subsection{Zero temperature: quantum-entropic rigidity.}

Let us first  evaluate the free energy Eq. (\ref{deltaf}) at zero temperature. 
It reduces to $\Delta_0$ and
this becomes with logarithmic  accuracy:

\begin{eqnarray}
&&\Delta_0=\frac{T^{3}\rho}{{\pi}^{2}\sqrt{\Sigma B}}\displaystyle\int_{0}^{\nu_s}
d y \int_{0}^{\nu_g}d x\;\left(\frac{\sqrt{x^2+y^2}}{2}-\frac{y}{2}\right)\nonumber\\
&&\approx\displaystyle\frac{T^{3}\rho}{{\pi}^{2}\sqrt{\Sigma B}}
\displaystyle\int_{0}^{\nu_g}d x\;\int_{\nu_g}^{\nu_s}d y\;\frac{x^2}{4y}=
\frac{T^{3}\rho}{{\pi}^{2}\sqrt{\Sigma B}}\frac{\nu_g^3}{12}
\ln{\left\{\frac{\nu_s}{\nu_g}\right\}}
\equiv\frac{Q_g^3 B}{12\pi^2
\sqrt{\rho\Sigma}}\ln{\left\{\frac{Q_s\sqrt{\Sigma}}{Q_g\sqrt{B}}\right\}}
\label{delta0}
\end{eqnarray}

\noindent
This result illustrates the crucial observation in \cite{jan} that the dominating 
contribution to $\Delta_0$ 
originates in the {\em long wavelength} on-string quantum fluctuations. The logarithm 
lives at
the lower limit $\nu_g$ of the $y$ integration, associated 
with the long wavelength 
on-string fluctuations with momentum  $q_y(y=\nu_g)\equiv
\nu_g T\sqrt{\rho}/\sqrt{\Sigma}\sim Q_g \sqrt{B/\Sigma}$. At
the same time, however, this free energy is dominated by large 
interstring momenta, $q_x(x=\nu_g)\equiv \nu_g T\sqrt{\rho}/\sqrt{B} \sim Q_g$,
and in this regard it is a short wavelength physics, as in the high temperature 
regime (next subsection). A more
careful evaluation of the integral in Eq. (\ref{delta0}) yields,

\begin{eqnarray}
\Delta_0=\displaystyle\frac{Q_{g}^{3}B}{24\pi^{2}\sqrt{\rho\Sigma}}
\left[2\ln{\left\{\frac{2Q_{s}\sqrt{\Sigma}}{Q_{g}\sqrt{B}}\right\}}+
\frac{5}{3}\right]
\label{delta01}
\end{eqnarray}

\noindent
Substituting the estimate Eq. (\ref{delta0}) in the
self-consistency equation (\ref{b}) we find the following equation for the
induced rigidity $B$:

\begin{eqnarray}
\frac{B}{d^2}=\displaystyle\frac{\partial^2}{\partial d^2} \left[\frac{Q_g^3 B}
{24\pi^2\sqrt{\rho\Sigma}}
\ln{\left\{\frac{Q_s^2 \Sigma}{Q_g^2 B}\right\}}\right]=-\displaystyle
\frac{\partial^2}{\partial d^2}\left[\frac{\pi c}{24\eta^3\Sigma_c}
\left(\frac{B}{d^2}\right)
\ln{\left\{d\left(\frac{B}{d^2}\right)\frac{a^2}{\eta^2
\Sigma_c}\right\}}\right]\;.
\label{self0}
\end{eqnarray}

\noindent 
Taking as an Ansatz for  $B(d)/d^2
\equiv \exp{[-\Phi(d)]}$, Eq. (\ref{self0}) can be solved with  exponential accuracy,
(quasi-classical approximation) which is valid when $\mu_g$ is small. It follows that 
\cite{jan}:

\begin{eqnarray}
B=Ad^2\exp{\left\{-\eta\left(\frac{54}{\pi}\right)^{1/3}\frac{1}{\mu_s^{1/3}}
\left(\frac{d}{a}\right)^{2/3}\right\}}=Ad^2\exp{\left\{-\eta\left(\frac{54}{\pi}
\right)
^{1/3}\frac{1}{\mu_g^{1/3}}\right\}}
\label{bofd0}
\end{eqnarray}

\noindent
Hence, instead of $B \sim exp ( - A / \mu_g)$ as followed from the ``naive'' 
collision-argument of section II
(Eq. \ref{qlc}), the Helfrich
method yields a stretched-exponential dependence on the coupling constant $\mu_g$. 
This stretched exponential
comes from the logarithm in Eq. (\ref{delta01}); upon neglecting this log one recovers 
the full exponential.
This logarithm finds in turn its origin in the long wavelength on-string fluctuations 
which are particularly dangerous
for the zero temperature elastic strings, reflecting the algebraic order of a single 
string. In the 
collision-language there is no room for these on-string long-wavelength fluctuations 
and the Helfrich
method suggests therefore a qualitatively different physics behind the induced quantum 
modulus. This seems
now confirmed by a numerical simulation. According to the numerical work of Nishiyama 
\cite{nishi}, 
$B \sim exp (- A' d^\beta)$
with $\beta = 0.808(1)$,  very close to our prediction $\beta = 2/3$ and very 
different from the naive
expectation $\beta =2$. We believe that the small difference between the numerical 
result and our result is
due to `fluctuation' corrections; Helfrich's method has the structure of a mean-field 
theory and it should be
possible to construct a perturbation expansion based on the difference between the 
exact action     
Eq. (\ref{partition}) and the `saddle point' action Eq. (\ref{action}).

\subsection{High temperature classical regime: $\nu_g \ll \nu_s \ll 1$}  

Consider now the high-temperature, fully classical regime $\nu_g \ll \nu_s \ll 1$, 
where
all mode frequencies are small compared to temperature. In this limit the integral
$I$ in Eq. (\ref{I}) becomes:

\begin{eqnarray}
&&I(\nu_s,\nu_g)=\displaystyle\int_{0}^{\nu_s}
d y \int_{0}^{\nu_g}d x\;\ln{\left(\frac{1-\exp{\{-\sqrt{x^2 +
y^2}\}}}{1-\exp{\{-y\}}} \right)}\nonumber \\
&&\approx -\displaystyle\int_{0}^{\nu_s}
d y \int_{0}^{\nu_g}d x\;\left(\frac{\sqrt{x^2+y^2}}{2}-\frac{y}{2} \right)+
\tilde{I} (\nu_s,\nu_g )\; ,
\label{Ih}
\end{eqnarray}

\noindent
where,

\begin{eqnarray}
\tilde{I}(\nu_s,\nu_g)\approx\displaystyle\int_{0}^{\nu_s}
d y \int_{0}^{\nu_g}d x\;\ln{\left(\frac{\sqrt{x^2 +y^2}}{y} \right)}\;.
\label{I1}
\end{eqnarray}

\noindent
Thus, from Eqs. (\ref{Ih}) and (\ref{deltaf}) it is obvious that at
high temperature, such that $\nu_g \ll \nu_s \ll 1$, the $I$-term 
splits into two parts with the opposite signs. The {\it{negative}} part of the 
integral $I$ exactly cancells the $\Delta_0$ part in the free energy $\Delta F$, 
in (\ref{deltaf}), while the positive part $\tilde{I}(\nu_s,\nu_g)$ plays the role of 
the high-temperature, "classical" limit of the free energy: 

\begin{eqnarray}
&&\Delta F(B)/L^2 \approx \displaystyle\frac{T^{3}\rho}{{\pi}^{2}\sqrt{\Sigma B}}
\tilde{I} (\nu_s,\nu_g )
\label{deltafh}
\end{eqnarray}

\noindent
The integration over $y$ in the integral in Eq. (\ref{I1}) is dominated by 
the interval  $y<\nu_g \ll \nu_s$.  Therefore, the upper integration limit over
$y$ could be made infinite with minor mistake:
$\nu_s \rightarrow \infty$. After that, the integral is made dimensionless by a 
change of
integration variables: $x\rightarrow \nu_g x'$ and $y\rightarrow \nu_g y'$. 
In this
way, $\nu_g$ can be scaled out from the double integral and we find,

\begin{eqnarray}
\tilde{I}=const\times \nu_g^{2}\; .
\label{I11}
\end{eqnarray}

\noindent
The $const$ can be calculated exactly to be equal to $\pi/4$.
Substituting Eq. (\ref{I11}) into Eq. (\ref{deltafh}) we find,

\begin{eqnarray}
\Delta F(B)/L^2\approx\frac{\pi T\sqrt{B}}{4d^{2}\eta^2 \sqrt{\Sigma}}=
\frac{\pi T\sqrt{B}}{4d^{3/2}\eta^2 \sqrt{\Sigma_c}}
\label{F2}
\end{eqnarray}

\noindent 
Substituting this result into the self-consistency equation
(\ref{b}) yields,

\begin{eqnarray}
\frac{B}{d^2}=\displaystyle\frac{\partial^2}{\partial d^2} \left[
\frac{\pi T\sqrt{B}}{4d^{3/2}\eta^2 \sqrt{\Sigma_c}} \right]\;.
\label{selft}
\end{eqnarray}

\noindent
Using as an Ansatz for the unknown function $B(d)=Cd^{\alpha}$, one finds 
from Eq. (\ref{selft}),

\begin{eqnarray}
B=\frac{9\pi^{2}T^{2}}{\Sigma_c d^3}
\label{B}
\end{eqnarray}

\noindent 
To be valid it  should be checked if this result is consistent with the initial 
assumption $\nu_g\ll 1$.
Substituting (\ref{B}) in the definition of $\nu_g$ and demanding that $\nu_g\ll 1$ we 
find,

\begin{eqnarray}
\nu_g\equiv\omega_g/T=(\pi/\eta d)\sqrt{B}/(T\sqrt{\rho})=
\frac{3\pi^2}{\eta^3}\mu_g\sim\mu_g\ll 1
\label{check}
\end{eqnarray}

\noindent
which is indeed consistent with the general condition that $\mu_g\ll 1$ in 
(\ref{dimdef}).

The solution (\ref{B}) can actually be directly checked using the fact that the high 
temperature
limit of the directed string gas in 2+1D is actually the same problem as the 
zero-temperature 
system of hard-core bosons in 1+1D. As discussed in \cite{jan}, the latter can be 
considered as
the compactified version of the former, where temperature plays the role of wrapping 
up one
of the dimensions.
The 1+1D hard core boson problem can be mapped on the trivial problem of 
non-interacting 
spinless fermions in 1+1D. Rewriting Eq. (\ref{B}) in terms of the dimensionful 
quantities characterizing 
the 1+1D hard core boson gas (mass of the bosons $M$, interparticle distance $d$) 
one finds\cite{jan},

\begin{eqnarray} 
B_{1d}(d)=\frac{9\pi^{2}}{\eta^4}\frac{\hbar^2}{M d^3}\;.
\label{check1}
\end{eqnarray}

\noindent  
Comparing this result with the exact results following from the spinless fermions one 
finds
that\cite{jan} $\eta =\sqrt{6}$, indicating that the Helfrich self-consistent phonon 
method is
remarkably accurate.

\subsection{Approaching the renormalized classical regime from the high temperature 
side.}

In the previous two subsections we demonstrated that the Helfrich method was 
completely
tractable analytically both in the zero- and high temperature limit. This is a lucky
circumstance because the equations are in principle quite cumbersome and the 
calculations only
become straightforward because we could exploit the smallness of $\mu_g$. 
This becomes different in the intermediate temperature regime. Despite
the expectation that also in this renormalized classical regime the answer is simple 
and 
actually the same as in the high temperature regime (Section II), we failed to find a 
closed analytic solution for the induced modulus at intermediate temperatures. We 
recall 
that the renormalized classical regime is defined as $\nu_g \ll 1\ll\nu_s$ which means 
that
the on-string Debye temperature is now large as compared to temperature. The largeness 
of $\nu_s$ 
makes it impossible to expand the integral $I$ in Eq. (\ref{deltaf}) in a way it
was done in Eq. (\ref{Ih}), because now the integration variable $y$ is not $\ll 1$ 
in the region $1<y\ll \nu_s$. To put it another way, the
difference between the two cases is due to the quantum zero-point contributions, 
because when $\nu_s \gg 1$ the on-string fluctuations have entered the quantum 
regime. Nevertheless, guided by the discussion 
in Section II (see text after Eq. (\ref{rcmeander})), we calculated numerically the 
ratio of the integral
$I$ to the "high temperature" expression $\tilde{I}$ in Eq. (\ref{I11}) in 
the "renormalized 
classical" region of parameters $\nu_g \ll 1\ll\nu_s$, see Fig. 1. As one might
expect, the ratio remains close to 1 and nearly constant in the wide interval of the
values of the parameter $\nu_s$ at the two fixed values of $\nu_g\ll 1$: $\nu_g = 0.02$;
$ 0.04$. 
Namely, as is 
seen in the 
Fig. 1, the interval of $\nu_s$'s starts in the "classical" region: 
$\nu_g\ll\nu_s < 0.5$ and 
goes deep into the "renormalized classical" region: $\nu_g\ll 1\ll \nu_s \approx 20$.
Nevertheless, the ratio $I/\tilde{I}$ is $\approx 0.95$ ($\nu_g=0.04$) or $\approx
0.96$ ($\nu_g=0.02$) and changes by less than $2\%$ of
its value in the interval $0.5\leq \nu_s\leq 20$.
Having this numerical evidence, we conclude that, within a numerical factor which is 
very close to  unity, the high-temperature solution for the induced rigidity $B$ Eq.
(\ref{B}) remains valid in the renormalized classical regime.

\subsection{Approaching the renormalized classical regime from low temperatures.}

As we already emphasized, it is a most crucial issue if the zero-temperature 
quantum-kinetic interactions survive at a finite temperature. Using the 
collision arguments, we already argued in Section II that this should be the
case. However, we also learned that these arguments fail qualitatively at
zero temperature. At the same time, as we discussed in subsection IV A the
Helfrich method indicates that the on-string long wavelength fluctuations
are decisive for the zero-temperature rigidity -- on general
grounds one expects that any quantum phenomenon related to long wavelength should
be quite vulnerable to the finiteness of temperature and it is therefore a-priori 
unclear
what to expect at a small but finite temperature. Fortunately, it turns out that
the Helfrich method remains tractable in this regime of small but finite temperature 
and here we will derive a controlled  solution for the induced modulus in this regime.
It turns out that the zero temperature modulus is robust at finite temperatures, up
to a crossover temperature $T_0$ which is however different from the simple estimate
presented in Section II.

Let us consider the low-temperature 
corrections to the quantum result Eq. (\ref{delta01}). We assume that in this regime
the induced modulus is finite so that a regime exists where temperature can become 
small
as compared to the interstring Debye temperature. We will find that this assumption is
self-consistently satisfied. Hence, we are now dealing with the limit  
$1\ll\nu_g\ll\nu_s$ and in this case the
double-integral in Eq. (\ref{I}) can be rewritten as, in exponential
accuracy,

\begin{eqnarray}
&&I\equiv I(\nu_s,\nu_g)=\displaystyle\int_{0}^{\infty}
d y \int_{0}^{\nu_g}d x\;\ln{\left(\frac{1-\exp{\{-\sqrt{x^2 +
y^2}\}}}{1-\exp{\{-y\}}} \right)}\nonumber\\
&&\displaystyle\approx\int_{0}^{\infty}d y \int_{0}^{\infty}d x\;\ln{\left(
1-\exp{\{-\sqrt{x^2 +y^2}\}}\right)}-\nu_g\int_{0}^{\infty}d y\; 
\ln{\left(1-\exp{\{-y\}}\right)}\;.
\label{Ilt}
\end{eqnarray}

\noindent
Recalling the definition Eq. (\ref{deltaf}), this leads to the following 
estimate ($\zeta(3)=1.202$),

\begin{eqnarray}
\Delta_T\approx\displaystyle\frac{T^2Q_g\sqrt{\rho}}{\sqrt{\Sigma}}\frac{\pi}{12}
-\frac{T^3\rho}{\sqrt{\Sigma B}}\left(\frac{\zeta(3)}{2\pi}\right) \qquad
\mbox{if}\qquad 1\ll\nu_g \ll \nu_s\;.
\label{delowt}
\end{eqnarray}

\noindent
Combining this estimate  with Eq. (\ref{delta01}) the free energy
becomes in the low temperature limit,

\begin{eqnarray}
&&\Delta F(B)/L^2 = \displaystyle\frac{Q_{g}^{3}B}{24\pi^{2}\sqrt{\rho\Sigma}}
\left[2\ln{\left\{\frac{2Q_{s}\sqrt{\Sigma}}{Q_{g}\sqrt{B}}\right\}}+\frac{5}{3}\right]
+\displaystyle\frac{T^2Q_g\sqrt{\rho}}{\sqrt{\Sigma}}\frac{\pi}{12}
-\frac{T^3\rho}{\sqrt{\Sigma B}}\left(\frac{\zeta(3)}{2\pi}\right)\;.
\label{deltaflt}
\end{eqnarray}

\noindent
It is directly clear that at sufficiently low temperatures the temperature dependent corrections
can be neglected in the self-consistency differential equation, which means that the quantum modulus
Eq. (\ref{delta01}) remains valid at the lowest temperatures. Hence, there has to be a temperature 
$T_0$ where the thermal fluctuations become of similar importance as the quantum fluctuations. This
$T_0$ can be estimated from Eq. (\ref{deltaflt}) by equating the leading temperature dependent 
correction $\sim T^2$ to the zero temperature term, 

\begin{eqnarray}
\displaystyle\frac{Q_{g}^{3}B}{24\pi^{2}\sqrt{\rho\Sigma}}
\left[2\ln{\left\{\frac{2Q_{s}\sqrt{\Sigma}}{Q_{g}\sqrt{B}}\right\}}+\frac{5}{3}\right]
\sim \displaystyle\frac{T^2Q_g\sqrt{\rho}}{\sqrt{\Sigma}}\frac{\pi}{12}\;.
\label{crosscnd}
\end{eqnarray}

\noindent
Substituting in Eq. (\ref{crosscnd}) the zero-temperature result for $B$,
Eq. (\ref{bofd0}),  one finds,

\begin{eqnarray}
T_0\propto \displaystyle \sqrt{B}\sim \exp{\left\{-\eta\left(\frac{54}{\pi}\right)^{1/3}
\frac{1}{2\mu_s^{1/3}}
\left(\frac{d}{a}\right)^{2/3}\right\}} =\exp{\left\{-\eta\left(\frac{54}{\pi}\right)
^{1/3}\frac{1}{2\mu_g^{1/3}}\right\}}
\label{kostt}
\end{eqnarray}

\noindent
Hence, we find that the elementary consideration in section II leading to the estimate 
Eq. (\ref{tcross11}) is in essence correct, except that the stretched exponential 
result has
to be used for the zero temperature modulus. Of course, $T_0$ has the same status as 
$\omega_g$ and a regime of small but finite temperatures exists where $\nu_g \gg 1$
due to the zero temperature quantum fluctuations. The physics behind this result can be 
deduced 
from the low temperature expansion Eq.(\ref{deltaflt}). In contrast to the 
renormalized
classical regime entered at $T > T_0$, the string gas Debye temperature $\omega_g$ 
acts
according to the expectations. For $T < \omega_g$ the fraction of the modes which
are thermally occupied diminishes quickly, making it possible to arrive at the simple
expansion Eq. (\ref{deltaflt}). At the same time, it follows from the form of this 
expansion
that the long wavelength aspect of the zero-temperature modulus is not as simple as 
discussed
in the beginning of this subsection. Eq. (\ref{deltaflt}) still starts out with the 
unmodified
zero temperature result despite the fact that temperature is finite and this term 
should be
destroyed immediately if the argument that the long wavelength on-string fluctuations 
immediately
loose their quantum character for any finite temperature would be correct. The 
resolution of
this apparent paradox is that the free energy is that of the effective  2+1D medium and the 
zero-temperature logarithm appears in the final answer, after integrating both 
interstring and
intrastring momenta. This implies a quantum rigidity for the system as a whole, and 
this in turn
leads to a diminishing of all thermal fluctuations, including those acting along the 
strings.
The implication  is that there is still a length scale where the quantum-induced 
modulus appears and when temperature 
becomes low enough the wavelength of the typical thermal fluctuations becomes large as 
compared to this
length scale and thereby inconsequential for the induced modulus. In this sense the 
basic argument of
section II survives in this self-consistent phonon language.  
The $T_0$ {\it{vs}} $d$ dependence as expressed in Eq. (\ref{kostt}) is plotted in
Fig. 2. Based on the observation made in Section II, that the dislocations can only 
proliferate at a Kosterlitz-Thouless temperature $T_{KT} \sim T_0$, we consider 
Fig. 2 as the phase diagram of the directed hard-core string-gas. 
$T_{0}(d)$ plays a role of the solidification line, which separates the 
low-temperature long-range ordered state of the string-system at $T<T_{0}(d)$ from 
the disordered string-gas state at the "high" temperatures: $T>T_{0}(d)$.

\section{Conclusions}
In this paper we have presented results of a detailed study of a gas of elastic
quantum strings in $2+1$ dimensions, interacting via a hard-core condition. 
The model mimics some essential features of a dynamical
stripe state, arising in the underdoped cuprate superconductors. From a more
general perspective, it relates to the theme of entropic interactions at finite
temperatures and quantum-kinetic interactions at zero temperature. 

Although in detail quite different, the physics at zero temperature falls in the
same category as for instance superexchange which is for good reasons also called
kinetic exchange. Physics which is associated with  kinetic energy (electron hops) 
at short distances becomes physics associated with order and potential energy at
long distances (antiferromagnetism). In the string gas, the short distance physics
corresponds with  the string fluctuations and the long distance physics is that of 
long range
order, breaking translational symmetry. As was emphasized in \cite{jan}, the elastic
string gas in 2+1D is a close relative of the Luttinger liquid in 1+1D. The same
order-out-of-disorder mechanism which renders the algebraic order of the hard core
bose gas in 1+1D to be generic, causes the true long range order in the string gas
at zero temperature.

Here we focussed mainly on the finite temperature physics of the quantum string gas.
We found this to be a rather non-trivial affair. The same basic mechanism which is 
responsible
for the zero-temperature kinetic interactions is responsible for the entropic 
interactions
at high temperature. We focussed on the question, how do these two regimes connect?

Against our initial expectations we found out that quantum-mechanics takes over rather
suddenly. We pointed out that there is a temperature scale below which the 
quantum-mechnical
nature of the string fluctuations becomes important. However, we found that these 
quantum
fluctuations are initially completely hidden: the entropic interactions continue to 
behave
as if the strings are fully classical. The reason is that the typical 
string-fluctuations 
responsible for the induced modulus live at frequencies which also decrease with 
temperature,
in such a way that the quantum cut-off stays effectively at infinity. As a result, 
the entropic
interactions stay in this `renormalized classical regime' in fact unrenormalized.

One could now have the impression that thermal fluctuations would continue to overrule
the quantum fluctuations down to the lowest temperatures. An additional motive could
be that the zero-temperature `stretched exponential modulus' which has been confirmed 
by
numerical simulations is due to long-wavelength on-string quantum fluctuations. One 
would
anticipate that these long-wavelength quantum fluctuations would be extremely 
vulnerable
to finite temperature. However, we found that the zero temperature kinetic 
interactions
are self-protecting in the regime of small but finite temperatures. Below the scale 
$T_0$, which is 
set by the zero-temperature modulus, quantum-mechanics starts to play a conventional 
role. Below
$T_0$ the phase space for thermal fluctuations shrinks rapidly and thereby also the 
influence
of the thermal fluctuations on the induced modulus. Hence, while this conventional 
intuition
failed badly at intermediate temperatures, it is correct at temperatures less than 
$T_0$.

Of course, the above picture rests entirely on the self-consistent phonon method 
invented
a long time ago by Helfrich. This approximate method is put here by us to the test in 
an
unprecedented way. However, we have confidence that the above conclusions are 
trustworthy.
After all, the hardest part is zero temperature where according to the Helfrich method 
a truely
novel mechanism is at work, giving rise to the stretched exponential. Except for some
small correction, likely due to `fluctuations around the mean-field', this stretched 
exponential turns out to be correct. Given that the high temperature limit is also
described rather accurately, it has to be that the intermediate temperature regime 
is also described accurately.

\section*{ACKNOWLEDGEMENTS}

This research was supported in part  by the Foundation
of Fundamental Research on Matter (FOM), which is sponsored by the
Netherlands Organization of Pure research (NWO).

\newpage
\begin{figure}
 \vbox to 4.0cm {\vss\hbox to -5.0cm
 {\hss\
       {\includegraphics{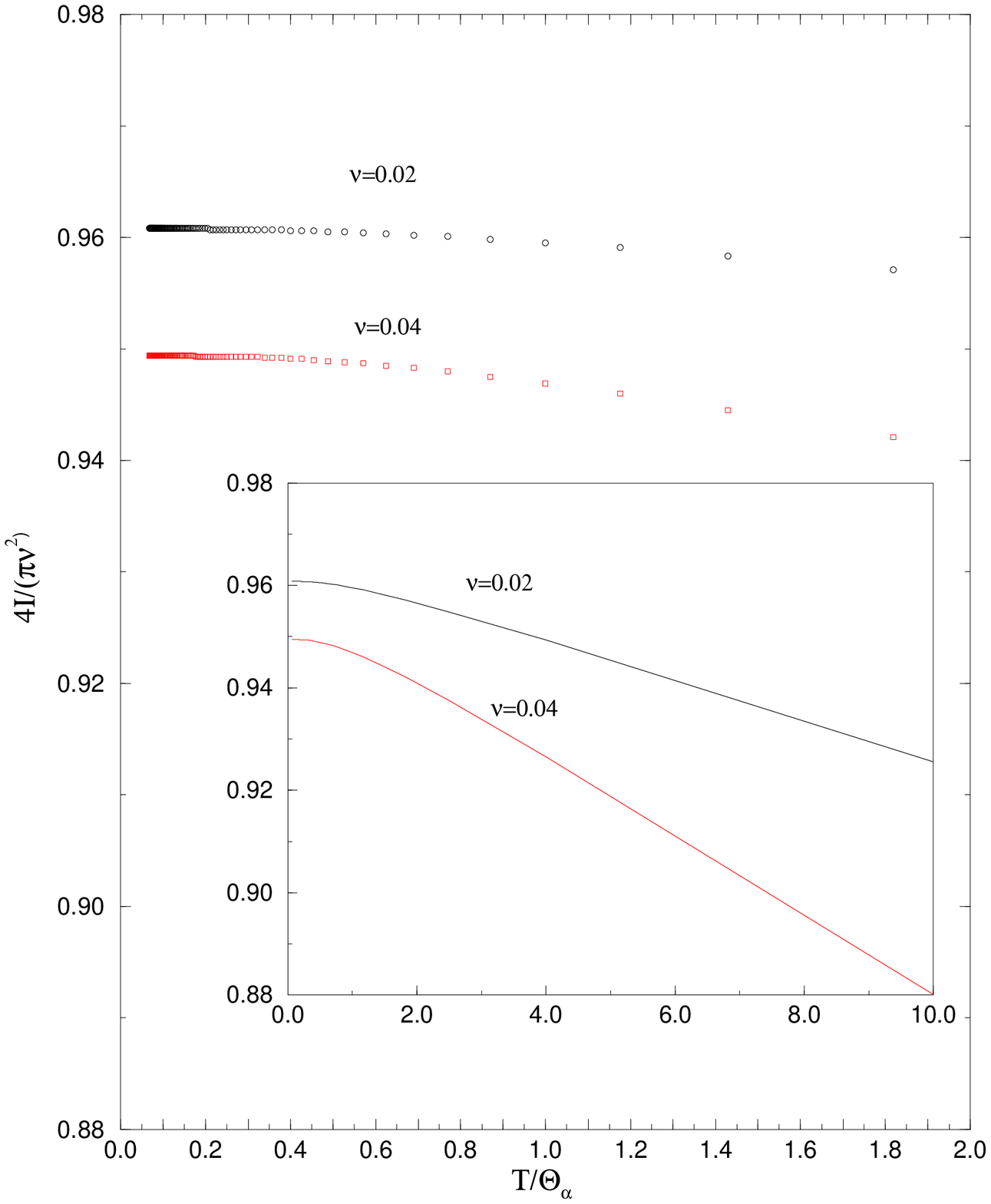}
       }
  \hss}
 }
\vspace{17cm}
\caption{Dependence of $I/(\pi\nu_g^2/4)$ on $\nu_s^{-1}=T/\omega_s$ for two 
different values of $\nu_g$: $\nu_g=0.02$ and $\nu_g=0.04$.} 
\label{fig1}
\end{figure}

\begin{figure}
 \vbox to 4.0cm {\vss\hbox to -5.0cm
 {\hss\
       {\includegraphics{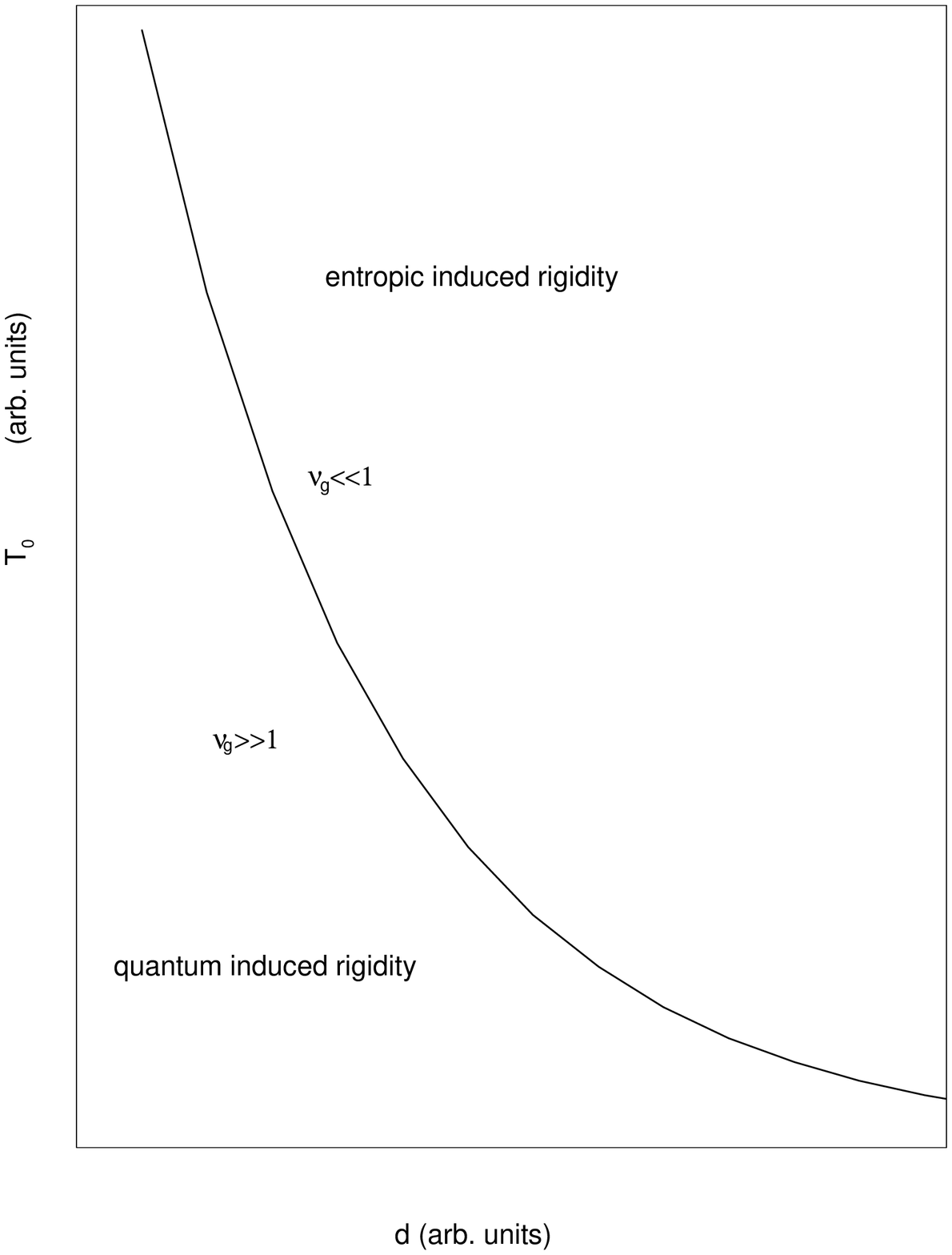}
       }
  \hss}
 }
\vspace{17.5cm}
\caption{Phase diagram of the directed hard-core string-gas: a solidification
temperature $T_0$ as function of the average interstring distance $d$.} 
\label{fig2}
\end{figure}

\begin{references}
\bibitem{finT}
The classical, finite temperature version of this problem has been
analyzed thoroughly in the past (see \cite{copper} and ref.'s therein) and
considered in the context of stripes in \cite{zahovs}.
\end{references}
\end{document}